\documentclass[twocolumn,english,superscriptaddress,secnumarabic,amssymb,amsmath,nobibnotes,aps,prd,showpacs]{revtex4}
\usepackage[latin1]{inputenc}
\usepackage{graphicx}
\usepackage{amssymb}
\usepackage{color}
\usepackage{float}
\usepackage{amsmath}
\usepackage{amsfonts}
\usepackage{dcolumn}
\usepackage{hyperref}
\usepackage{amsthm}
\usepackage{color}

\begin{document}
 
\title{Dynamics and chaos in the unified scalar field cosmology II. System in a finite box}

\author{Giovanni Acquaviva}
\email{gioacqua@utf.troja.mff.cuni.cz}
\affiliation{Institute of Theoretical Physics, Faculty of Mathematics and Physics, 
 Charles University in Prague, 18000 Prague, Czech Republic}

\author{Georgios Lukes-Gerakopoulos}
\email{gglukes@gmail.com}
\affiliation{Institute of Theoretical Physics, Faculty of Mathematics and Physics, 
 Charles University in Prague, 18000 Prague, Czech Republic}

\begin{abstract}
 We revisit the global dynamics of unified dark matter cosmological models and
 analyze it in a new dynamical system setting. In particular, by defining 
 a suitable set of variables we obtain a bounded variable space, a feature that
 allows a better control of the critical elements of the system. First, we give
 a comprehensive cosmological interpretation of the critical points. Then, we
 turn our focus on particular representative trajectories with physically
 motivated initial conditions studied in the first paper of the series,
 and we discuss how the scale factor relates to the equation of state parameter.
 We review and complement these results in the light of the new variable approach
 by discussing the issue whether the system is chaotic or not.
\end{abstract}


\maketitle

\section{Introduction}

The current standard cosmological model is the $\Lambda$ Cold Dark Matter
($\Lambda$CDM) scenario, which includes two elements whose components 
have unknown nature: the dark matter and the dark energy.  This simple model 
fits quite well the observed behavior of our universe \cite{ade2015}\cite{riess2016} 
at least from the matter-dominated epoch, where structures formation starts.  
However, the nature of both the dark components is still elusive and it is matter 
of intense debate, see, e.g., \cite{li2013,patr2016} for a broad overview.  While part
of the literature approaches the study of this topic by keeping the dark sector
divided into two separate components, efforts have been put also into unifying
the effect of both dark energy and dark matter under the same mechanism, for
example through the modified Chaplygin gas models (see, e.g., \cite{wu2007} and
references therein). Along the same lines, in \cite{Varum00,Bertacca07} a scalar
field potential was introduced, which is able to act both as dark matter and dark energy, 
hence dubbed Unified Dark Matter (UDM).  Since its introduction it has been subject of detailed
analysis \cite{LGBC08,Bertacca08,BLG08,Bertacca10,Paliathanasis14,Starkov16}.\\

Here we revisit the dynamical aspect of the UDM cosmology first addressed
in \cite{LGBC08}, where such scalar field was studied in the background of a 
spatially curved Friedmann-Robertson-Walker (FRW) spacetime. The resulting
cosmological model was recast as a Hamiltonian system of two coupled
oscillators and studied as such. The phase space of the system was unbounded and
allowed trajectories to escape to infinity. Moreover, in the case of a positive
spatial curvature indications of chaotic behavior were found for some sets of
initial conditions. These indications were based on transient features of
the characteristic Lyapunov number and of a similar chaotic indicator, so
the chaotic behavior was called chaotic scattering instead of chaos. 
In general, when one addresses the issue of whether a dynamical system suffering
escapes exhibits chaos or not, Lyapunov-like indicators should be used with
caution, see, e.g., \cite{Contop99}.  In fact, in the literature there is a variety of 
approaches different from Lyapunov-like indicators to address the search for chaos in scalar field 
cosmologies, starting, e.g., from Page \cite{page1984} to more recent developments such as \cite{hrycyna,topo} and references therein.\\

In the present paper we mainly address the study of the same cosmological model 
as in \cite{LGBC08}, but we introduce a different parametrization of the dynamical
variables. This parametrization confines the system ``in a box'', i.e. in a
bounded variable space. Such reparametrization has been widely used in previous
works on cosmological dynamical systems (see, e.g., \cite{madsen1988,gol1999,campos2001,heinzle2005,vanhelst2002})
and it is especially useful in treating cosmological models with spatial positive
curvature, where the possibility of recollapsing universes has to be taken into 
account. In fact, in such models, the usual definition of 
{\it expansion-normalized variables} is doomed to present singular behavior in
the turning points of the scale factor.\\

Such reparametrization allows us to uncover new information regarding the
dynamics of the system. On one hand it allows us to study the specific
features of the cosmological model encoded in the critical points. On the
other hand it allows us to have a new perspective on the dynamics of the system.
Namely, in the box version of the UDM model we have sinks, and every orbit will
end up in one of these sinks. Thus, by studying the structure of the basins 
of attraction of these sinks we can tell whether the system is chaotic or not.
If a system is chaotic, then the sensitivity on initial condition will
be represented by a fractal structure of the basins of attraction, see, e.g.,
\cite{Kennedy91,Cornish97}, and references therein. Our analysis shows that
the basins of attraction have not a fractal structure, so the orbital dynamics
of the UDM model should not be chaotic, which is in fact 
in agreement with \cite{topo}.\\

The paper is structured as follows.  In Sec.~\ref{sec:TheAn} we present and
analyze the UDM cosmological model; we first briefly review the Hamiltonian
formulation and then provide the bounded dynamical system; the latter is analyzed
in detail from both mathematical and cosmological perspectives. 
In Sec.~\ref{sec:orbital} we focus on the analysis of some representative orbits
of the variable space, providing their interpretation in cosmological terms;
moreover, we address the problem of the lack of chaos in the system by
inspecting the basins of attraction of the sinks. Eventually,
Sec.~\ref{sec:concl} summarizes of the main results. In this work the geometric
units are employed.

\section{Theoretical analysis} \label{sec:TheAn}

\subsection{The cosmological model}\label{sec:cosmomod}

We consider a cosmological model described by the homogeneous and isotropic 
FRW metric
\begin{equation}\label{frw}
 ds^2 = - dt^2 + a^2(t) \left[ \frac{dr^2}{1-k r^2} + r^2 d\theta^2 + r^2 \sin^2\theta d\phi^2 \right] \quad,
\end{equation}
where $a$ is the scale factor.  Open, flat and closed spatial geometries correspond
respectively to $k<0, k=0$ and $k>0$.  In the following we will be interested in
the spatially closed case, which allows for recollapsing cosmological scenarios.
As a source in Einstein's field equations we consider the energy-momentum tensor
$T^{\mu\nu}$ of a massive scalar field, uniquely described by its energy density
and its pressure given respectively by
\begin{align}
 \rho_{\phi} &= -\, T^0_0 = \frac{1}{2}\dot{\phi}^2 + U(\phi)\label{ener} \quad, \\
 p_{\phi} &= T^i_j = \frac{1}{2}\dot{\phi}^2 - U(\phi)\, \quad,\label{press}
\end{align}
where $\dot{}$ indicates the derivative with respect to the cosmic time $t$.
Hereafter, we assume an effective barotropic equation of state (EoS) 
$p_{\phi}=w\, \rho_{\phi}$ for the scalar field. The potential $U$ in
\cite{Bertacca07,LGBC08} was chosen to have the form
\begin{equation}\label{pot}
 U(\phi) = c_1\, \cosh^2\left( \sqrt{\frac{3}{8}}\, \phi \right) + c_2\, \quad,
\end{equation}
where $c_1$ and $c_2$ are real constants.  In order for the field to have a real
non-negative mass, such constants are required to satisfy the conditions
$c_1\geq0$ and $c_1+c_2\geq0$ \cite{Bertacca07,LGBC08}. As already stated before,
this kind of scalar fluid has been considered as a possible candidate for
unifying the effect of the whole dark sector.\\

An alternative form of potential \cite{Paliathanasis14} reads
\begin{align}\label{pot2}
 U(\phi) &= b_1 \left(1+3\cosh^2\left( \sqrt{\frac{3}{8}}\, \phi \right)\right) \nonumber \\
         &+ b_2 \left(3 \cosh\left( \sqrt{\frac{3}{8}}\, \phi \right)+\cosh^3\left( \sqrt{\frac{3}{8}}\, \phi \right)\right) \quad.
\end{align}
In Appendix \ref{appen} we discuss briefly the case $b_1=b_2$, i.e.
\begin{align}\label{pot3}
 U(\phi) &= b_1 \left(1+\cosh\left( \sqrt{\frac{3}{8}}\, \phi \right)\right)^3 \quad,
\end{align}
which is a particular case of a potential already introduced in \cite{Varum00}.

By implementing metric \eqref{frw} and source \eqref{ener}, \eqref{press} into
Einstein's field equations and explicating the covariant conservation of
energy-momentum, the following system is obtained
\begin{align}
 H^2 + \frac{k}{a^2} &= \frac{1}{6}\dot{\phi}^2 + \frac{1}{3}\, U(\phi) \label{efe1} \quad,\\
 2\, \dot{H} + 3\, H^2 + \frac{k}{a^2} &= - \frac{1}{2}\dot{\phi}^2 + U(\phi)\label{efe2} \quad,\\
 \ddot{\phi} + 3\, H\, \dot{\phi} &+ \partial_{\phi} U = 0 \label{kg}\, \quad,
\end{align}
where we employed the Hubble expansion parameter $H=\dot{a}/a$.

\subsection{UDM as coupled oscillators} \label{sec:UDMHam}

In \cite{LGBC08} the UDM cosmology was expressed in terms of two coupled 
oscillators. To achieve that the variables were changed from $(a,\phi)$ to
\begin{eqnarray} 
      \label{trans}
      x=A ~a^{3/2} \sinh{(c ~\phi)} \nonumber \\
      y=A ~a^{3/2} \cosh{(c ~\phi)}
\end{eqnarray}
with $c^{2} =3/8$. This transformation results in a Hamiltonian function
\begin{eqnarray}
      \label{Hamxy}
      {\cal H} =\frac{1}{2}\left[(p_{x}^2-\frac{3}{4}c_2 x^2)-(p_{y}^2-\frac{3}{4}(c_1+c_2) y^2)\right] \nonumber \\
       -\frac{1}{2}\left[3^{4/3} k (y^2-x^2)^{1/3}\right]
\end{eqnarray}
where $ p_x=\dot{x} $, $ p_y=-\dot{y}$ denote the canonical momenta.
The Hamiltonian function~\eqref{Hamxy} describes two coupled oscillators, which makes
the UDM analysis more dynamically intuitive. However, since $c_1+c_2\geq0$,
the constant $c_2$ can be negative and one of the oscillators offers hyperbolic 
solutions; this implies the presence of trajectories escaping to infinity
$(a\rightarrow \infty)$ even 
if $k>0$.  Moreover, a closed universe implies also the possibility of recollapsing 
models $(a\rightarrow 0)$. Since in this coupled oscillators formalism the system 
suffers from escapes to infinity and recollapses, the phase space is
not bounded (see \cite{Starkov16} for details), and this brings along certain
difficulties in performing a dynamical study of the UDM cosmology \cite{LGBC08}.
In the following section we overcome the unboundedness by performing a novel
reparametrization.

\subsection{UDM in a box} \label{sec:UDMboxD}

In constructing a cosmological dynamical system with flat or negative curvature,
it is usually sufficient to define dimensionless variables by normalizing the
relevant dynamical quantities over the expansion $H$.  However, as we are are
considering geometries with $k>0$, such choice includes recollapsing solutions.  
While cosmological observations \cite{ade2015} tend to rule out 
the presence of a non-vanishing spatial curvature, we are especially interested in including
into the picture the possibility of recollapse because of its rich and interesting dynamics 
\footnote{Note that in the numerical analysis of Subsec. \ref{sec:orbital} 
we choose a value of curvature which is marginally compatible with the observational bounds
(as was done in \cite{LGBC08}).}.
However, a recollapsing scenario can lead to singular behaviors of the usual 
normalized variables in the turning points of the scale factor, where $H=0$.
This problem can be circumvented by noting that, while $H$ alone can vanish 
during the evolution, the left-hand side of eq.~\eqref{efe1} instead is always positive.
Hence, one can define a new set of variables by normalizing over the quantity
\begin{equation}
 D = \sqrt{H^2 + \frac{k}{a^2}}\, \quad. \label{D}
\end{equation}
The dimensionless variables obtained with such choice are the following
\begin{align}
 X_{\phi} &= \frac{\dot{\phi}}{\sqrt{6}\, D}\label{xphi} \quad,\\
 X_{U} &= \frac{\sqrt{U}}{\sqrt{3}\, D}\label{xu} \quad,\\
 X_H &= \frac{H}{D}\label{xh} \quad,\\
 X_{\partial U} &= - \frac{\partial_{\phi}U}{U}\label{xdu}\quad, \\
 F &= \frac{\partial^2_{\phi} U}{U} \label{xddu}\, \quad.
\end{align}
The definitions \eqref{xdu} and \eqref{xddu} are inspired by the quantities used
in the study of cosmological tracking solutions (see, e.g., \cite{stein1999,ng2001}).  
The function $F$ has to be included in order to close the system, but we stress
the fact that it does not act as an independent variable. In fact, knowing in
principle the dependence of the potential on the field, one could invert
eq.~\eqref{xdu} and obtain $\phi=\phi\left(X_{\partial U}\right)$, which could
then be plugged into eq.~\eqref{xddu} to obtain $F\left(X_{\partial U}\right)$.
In the specific case of potential~\eqref{pot}, one can see that the function 
$X_{\partial U}$ is invertible in the whole range of $\phi$ only if
$\displaystyle \alpha=c_2/c_1\geq-1/2$; however, if $-1<\alpha<-1/2$ the inversion
can be carried out only piecewise because the function is not monotonic anymore. 
Here, we focus our attention to the range $\alpha\geq -1/2$ which, apart
from including physically relevant situations, allows for a quite general
description, as the features of the critical elements of the variable space
turn out to be effectively independent of the specific value of $\alpha$.  
Hence, if $\alpha\geq -1/2$, the function $X_{\partial U}$ is monotonic in $\phi$
and bounded. It is then straightforward to find $\phi$ as function of $X_{\partial U}$, i.e.
\begin{equation} \label{fieldinv}
 \phi = \sqrt{\frac{2}{3}}\, \log \left[ \frac{\sqrt{1+\frac{8}{3}\, X_{\partial U}^2
 \alpha\, (1+\alpha)}-\sqrt{\frac{2}{3}}\, X_{\partial U}\, (1+2\alpha)}{1+\sqrt{\frac{2}{3}}\, X_{\partial U}} \right]\, 
\end{equation}
as a consequence of the boundedness of $X_{\partial U}$, which is defined in the
range $X_{\partial U}\in [-\sqrt{3/2},\sqrt{3/2}]$. Expression~\eqref{fieldinv} 
can be plugged into eq.~\eqref{xddu} in order to express it as a function of
$X_{\partial U}$, that is
\begin{equation}
 F\left( X_{\partial U} \right) = \frac{\left( 1 + 2\, \alpha \right)\,
 \sqrt{ 1 + \frac{8}{3}\, \alpha\, \left( 1+ \alpha \right)\, X_{\partial U}^2}
 - 1}{\frac{8}{3}\, \alpha\, \left( 1 + \alpha \right)}\, \quad.
\end{equation}\\

With regard to other variables~\eqref{xphi}-\eqref{xh}, it is easy to check that
they are bounded as well. First of all, by recasting eq.~\eqref{efe1} in terms
of the normalized variables one obtains
\begin{equation}
 X_{\phi}^2 + X_U^2 = 1\, \quad.\label{constr}
\end{equation}
Given that $X_U\geq0$, then $X_{\phi}\in [-1,1]$ and $X_U \in [0,1]$. Finally,
the variable $X_H$ is clearly defined in the interval $[-1,1]$; moreover, we
notice that this variable is positive/negative iff the universe is expanding/collapsing.\\

In order to build the dynamical system, it will be useful to write both 
eq.~\eqref{efe2} and the evolution equation of the quantity $D$ in terms of the
new variables
\begin{align}
 \frac{\dot{H}}{D^2} &= \frac{3}{2} \left( X_U^2 - X_{\phi}^2 \right) - X_H^2 - \frac{1}{2} \label{hdot} \quad,\\
 \frac{\dot{D}}{D^2} &= \frac{3}{2} X_H\, \left( X_U^2  - X_{\phi}^2 - 1 \right) \label{ddot} \quad.
\end{align}
Defining the new ``time'' derivative $X' = D^{-1}\dot{X}$, the autonomous
dynamical system is constructed by taking the prime derivative of the normalized
variables \eqref{xphi}-\eqref{xdu}. Then, by implementing eq.~\eqref{kg} and
eqs.~\eqref{constr}-\eqref{ddot}, the system takes the following form
\begin{align}
 X_{\phi}' &= \sqrt{\frac{3}{2}}\, \left( 1 - X_{\phi}^2 \right)\, \left( X_{\partial U} - \sqrt{6}\, X_H\, X_{\phi} \right)\label{phip} \quad,\\
 X_H' &= \left( 1 - X_H^2 \right)\, \left( 1 - 3\, X_{\phi}^2 \right)\label{hp} \quad, \\
 X_{\partial U}' &= -\sqrt{6}\, X_{\phi}\, \left( F - X_{\partial U}^2 \right)\label{dup} \quad,
\end{align}
where F is the function of $X_{\partial U}$ previously defined, which in terms
of the parameters of the potential reads
\begin{equation}
 F = \frac{\left( c_1 + 2\, c_2 \right)\, \sqrt{c_1^2 + \frac{8}{3}\, c_2\,
 \left( c_1 + c_2 \right)\, X_{\partial U}^2} - c_1^2}{\frac{8}{3}\, c_2\, \left( c_1 + c_2 \right)}\, \quad.
\end{equation}
By building the system in this way we decouple the evolution of $D$ from the rest
of the variables and, through the constraint~\eqref{constr}, we can
disregard the evolution of $X_U$.  Note that the quantities $X_{\partial U}^2$ 
and $F$ are known in cosmology as the {\it slow-roll parameters}, $\epsilon$ and 
$|\eta|$ respectively; this means that an exponential behavior of the scale factor 
is an expected feature whenever $|X_{\partial U}| \ll 1$ \footnote{An analogous 
condition on $F$ would be relevant for the primordial inflationary era, as it 
determines the duration, or $e$-folds, of the exponential expansion.}.
 
The scale factor as a function of the new variables is
 \begin{align} \label{scfnewvar}
  a=\frac{\sqrt{3 k}~ X_U}{\sqrt{1-{X_H}^2}\sqrt{c_1 \cosh^2{\left(\sqrt{\frac{3}{8}}\phi\right)}+c_2}} \quad, 
 \end{align}
where $\phi$ is given by the eq.~\eqref{fieldinv}.

\subsection{Analysis and interpretation of critical points} \label{sec:critical}

\begin{figure}[ht!!!]
 \begin{center}
 {\includegraphics[width=0.25\textwidth]{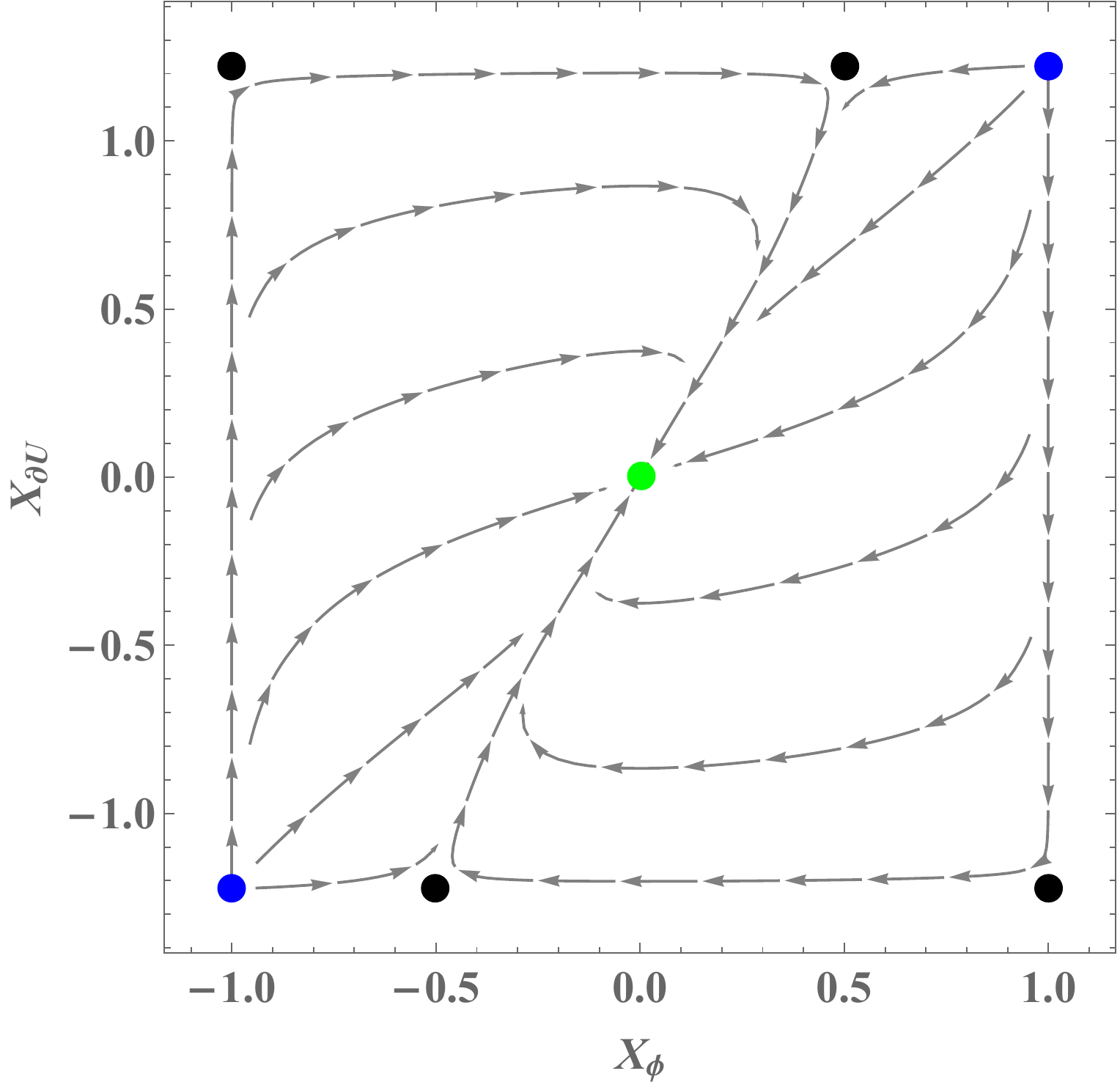}
 \includegraphics[width=0.25\textwidth]{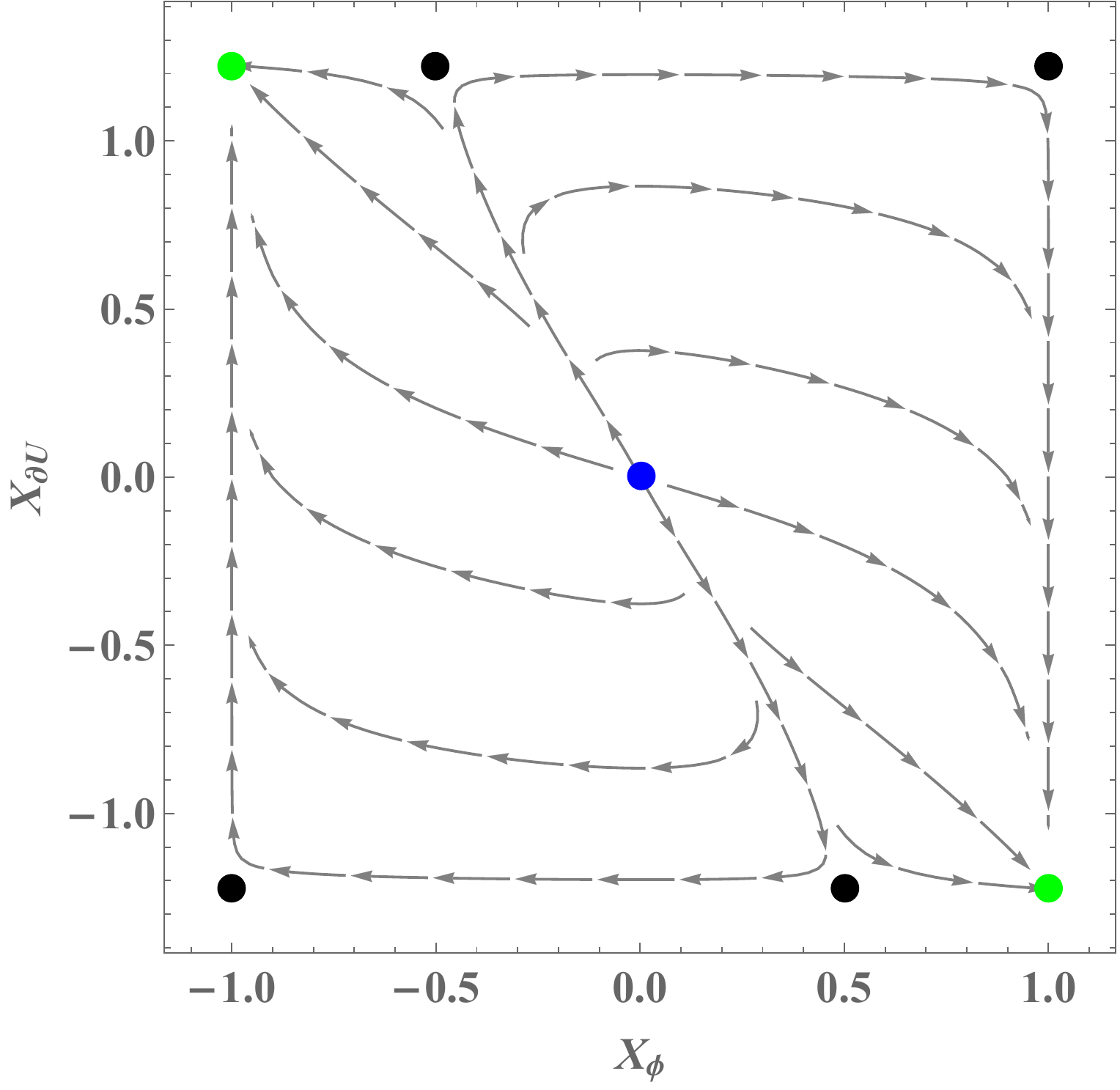}}
 \caption{\label{xh_inv}Invariant sets $X_H=1$ (left panel) and $X_H=-1$ (right
 panel).  The points identify sources (blue), saddles (black) and sinks (green).}
\end{center}
\end{figure}

\begin{figure}[ht!!!]
 \begin{center}
 {\includegraphics[width=0.25\textwidth]{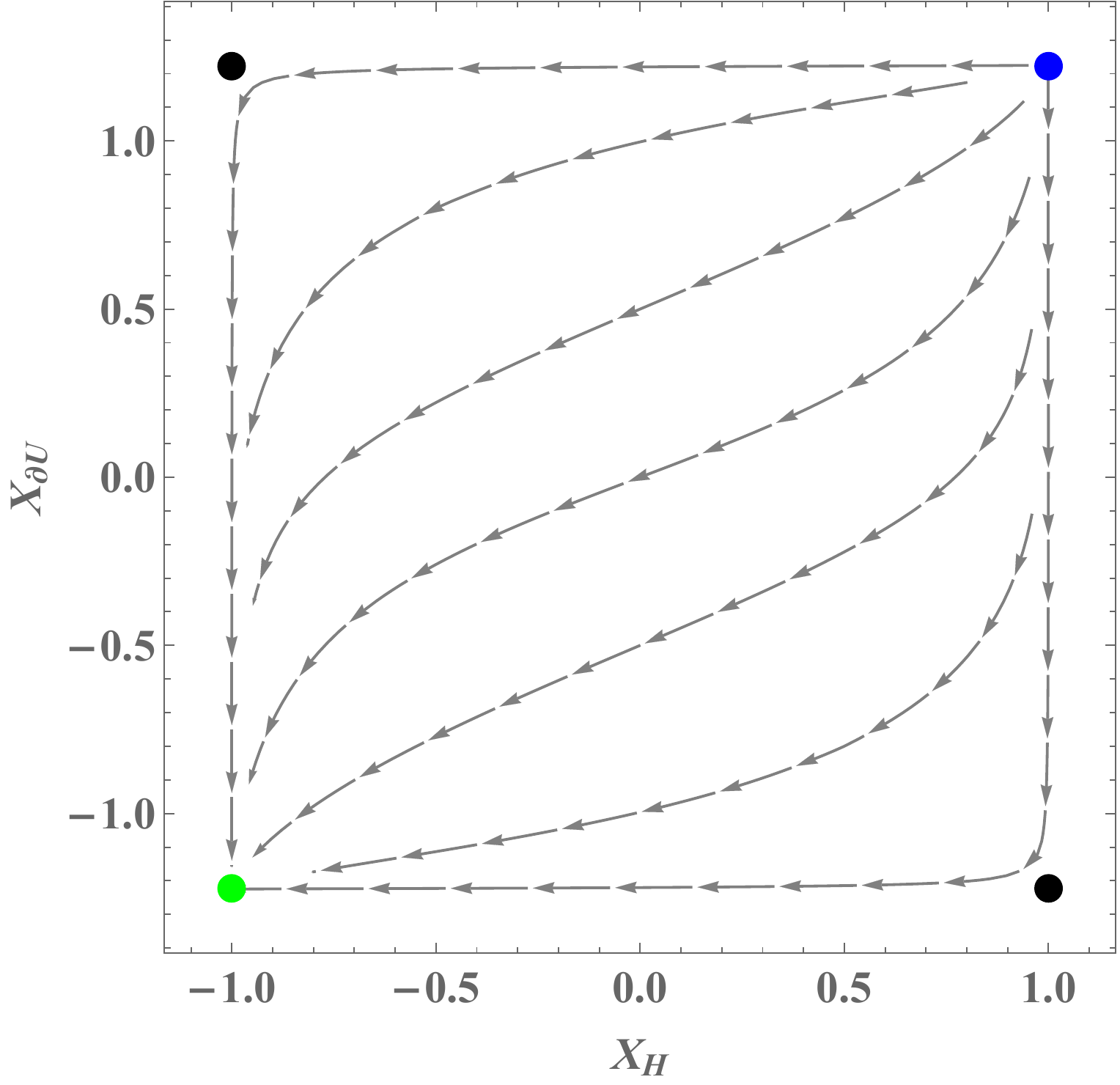}
 \includegraphics[width=0.25\textwidth]{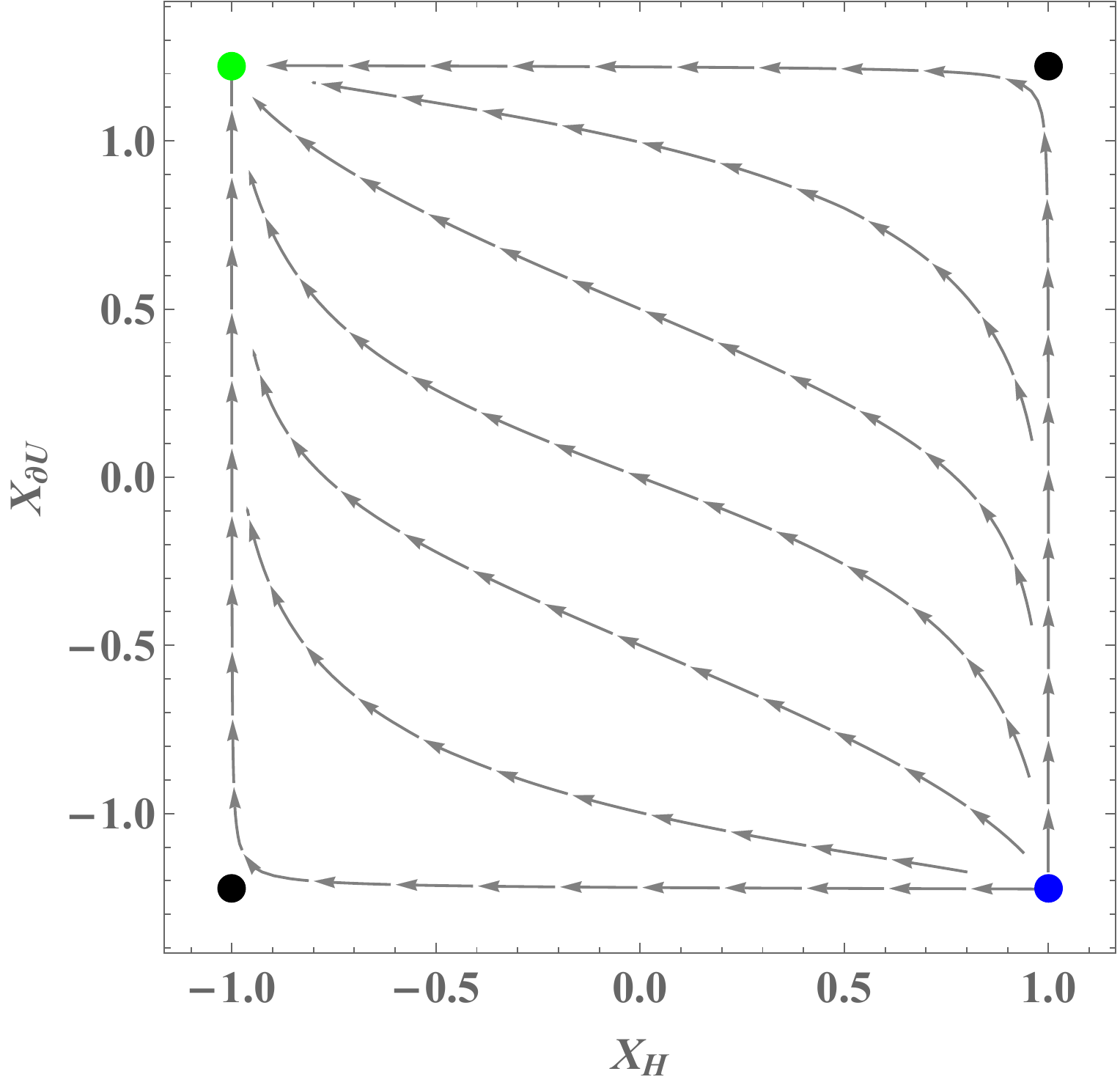}}
 \caption{\label{xphi_inv}Invariant sets $X_{\phi}=1$ (left panel) and 
 $X_{\phi}=-1$ (right panel).  The points identify sources (blue), saddles 
 (black) and sinks (green).}
\end{center}
\end{figure}

\begin{figure}[ht!!!]
 \begin{center}
 {\includegraphics[width=0.25\textwidth]{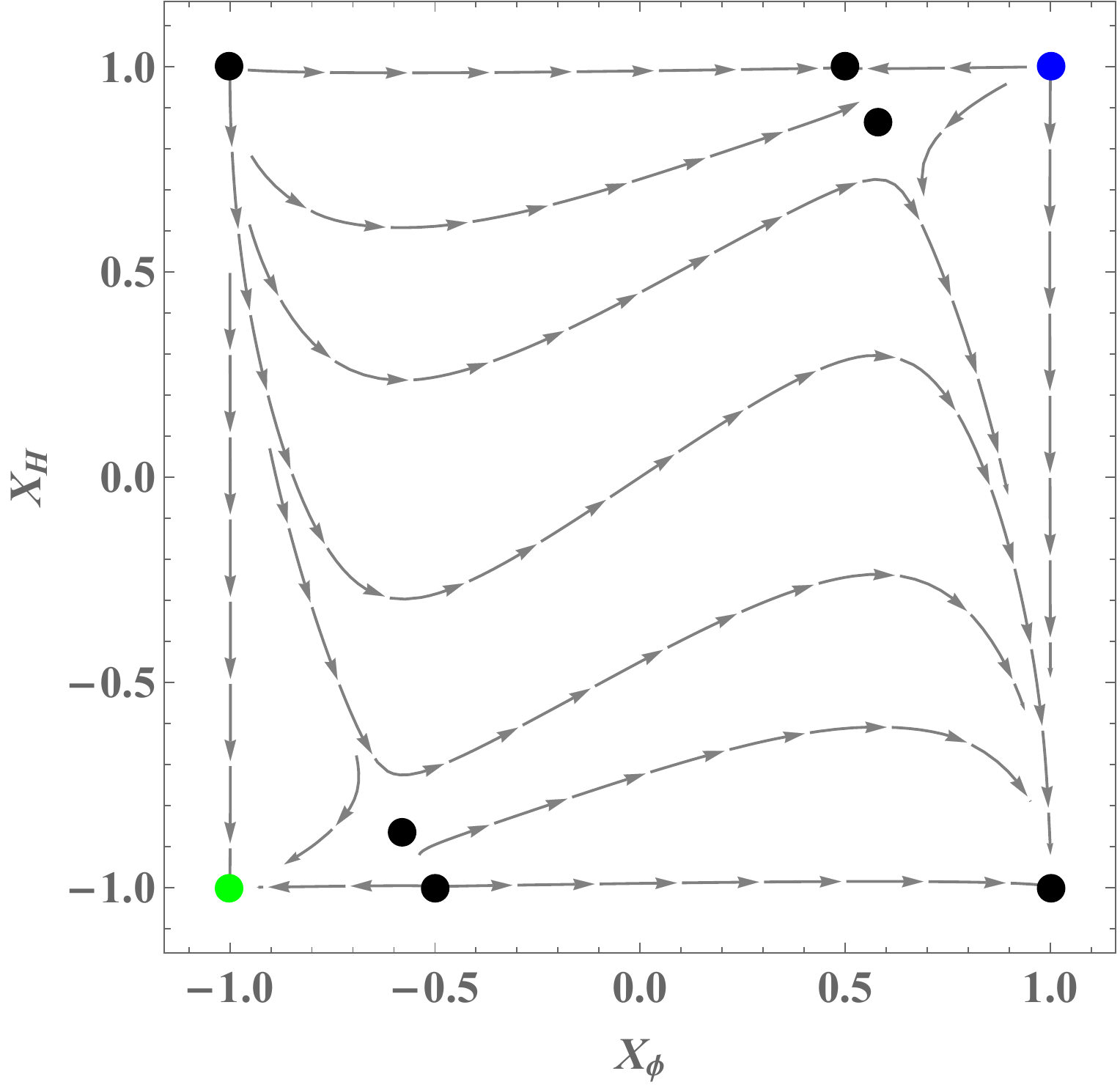}
 \includegraphics[width=0.25\textwidth]{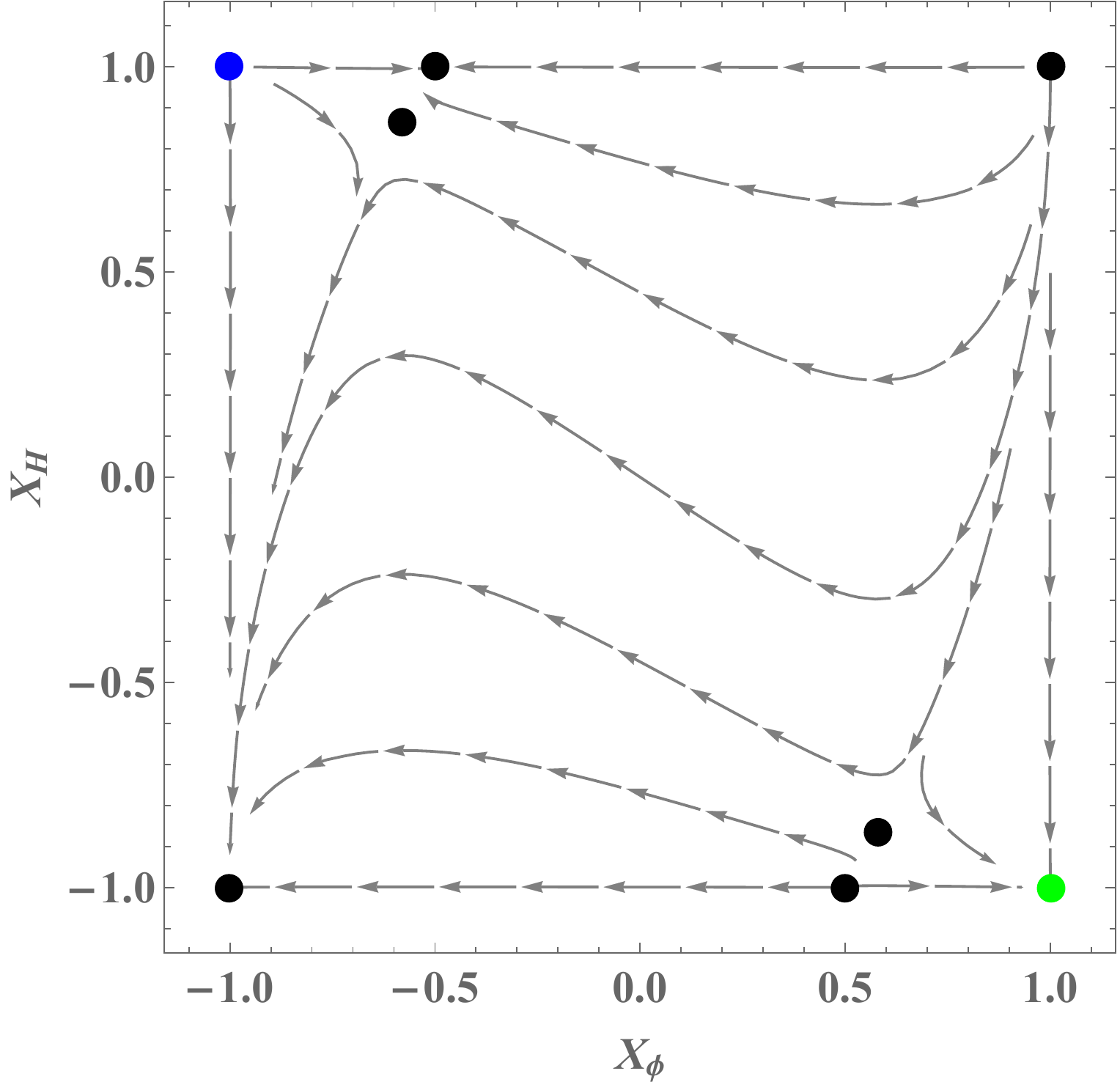}}
 \caption{\label{xdu_inv}Invariant sets $X_{\partial U}=\sqrt{3/2}$ (left panel)
 and $X_{\partial U}=-\sqrt{3/2}$ (right panel).  The points identify sources 
 (blue), saddles (black) and sinks (green).}
\end{center}
\end{figure}

The critical points of the system are those values of the variables
$\{ X_{\phi} , X_H , X_{\partial U} \}$ for which
${\bf X}' = 0$ is satisfied.  The stability of critical points is then analyzed
by inspecting the eigenvalues of the Jacobian matrix evaluated at the points
themselves. If the real part of all the eigenvalues is negative (resp. positive)
then the critical point is a sink (resp. a source), while if the sign of the
eigenvalues is mixed then the critical point is a saddle; in the latter case,
the stable eigendirections are given by the eigenvectors corresponding to the
negative eigenvalues.  For the system~\eqref{phip}-\eqref{dup} we list the 
critical points, their stability and cosmological interpretation in Table \ref{tab1}, 
while their location inside the invariant subsets are shown in 
Figs.\ref{xh_inv}-\ref{xdu_inv} for the specific choice $\alpha=1$.  
In such plots we specify with dots the locations of sources (blue), 
sinks (green) and saddles (black). It is worth noting that both the location and
the stability of the critical points are independent of the values of the
parameters $c_1$ and $c_2$ as long as their ratio $\alpha\geq-1/2$, so that the situations
portrayed are easily generalizable. Here we discuss in detail the cosmological
interpretation of the critical points. This can be done by calculating the
{\it deceleration parameter} $q$ and the {\it effective EoS parameter} $w$,
which are given respectively by
\begin{align}
 q &= \frac{3 X_{\phi}^2 - 1}{X_H^2} \quad,\\
 w &= 2\, X_{\phi}^2 - 1 \quad. \label{wEoS}
\end{align}
Eventually, one can calculate the cosmic time-dependent scale factor of the
models by integrating eqs.~\eqref{hdot},~\eqref{ddot} in the critical points.\\

\setlength{\extrarowheight}{10pt}
\setlength\tabcolsep{10pt}
\begin{table}
 \begin{center}
  \begin{tabular}{c||c|c|c}
    $ \{ X_{\phi} , X_H , X_{\partial U} \}$ & \ stability & $q$ & $w$ \\\hline
    $ \{ 0 , -1 , 0 \}$ & source & -1 & -1 \\
    $ \{ 0 , 1 , 0 \}$ & sink & -1 & -1 \\
    $ \{ \pm1 , 1 , \pm\sqrt{3/2} \}$ & source & 2 & 1 \\
    $\{ \pm1 , -1 , \mp\sqrt{3/2} \}$ & sink & 2 & 1 \\
    $\{ \pm1 , 1 , \mp\sqrt{3/2} \}$ & saddle & 2 & 1 \\
    $\{ \pm1 , -1 , \pm\sqrt{3/2} \}$ & saddle & 2 & 1 \\
    $\{ \pm1/2 , 1 , \pm\sqrt{3/2} \}$ & saddle & -1/4 & -1/2 \\
    $\{ \pm1/2 , -1 , \mp\sqrt{3/2} \}$ & saddle & -1/4 & -1/2 \\
    $\{ \pm1/\sqrt{3} , \pm\sqrt{3}/2 , \sqrt{3/2} \}$ & saddle & 0 & -1/3 \\
    $\{ \pm1/\sqrt{3} , \mp\sqrt{3}/2 , -\sqrt{3/2} \}$ & saddle & 0 & -1/3 \\
  \end{tabular}
 \end{center}
\caption{\label{tab1} Critical points of the system, their stability and 
corresponding cosmological parameters.}
\end{table}

There are three sources in the variable space. The point $\{ 0,-1,0 \}$ is
located in the invariant subset $X_H=-1$ and hence describes collapsing models
with negligible curvature with respect to the Hubble parameter. Moreover, the
fact that $X_{\phi}=0$ implies that the field's potential energy dominates
over the kinetic energy; hence, these models are collapsing exponentially,
i.e. $a \sim e^{-H_0\, t}$, starting from the infinite past of cosmic time.
As expected, the parameters $q$ and $w$ have the values corresponding to a 
cosmological constant.  The other two sources $\{ \pm1 , 1 , \pm\sqrt{3/2} \}$ 
instead, being in the invariant set $X_H=1$, describe expanding models with 
negligible curvature. In these points the kinetic energy dominates and the 
cosmological parameters are $q=2$ and $w=1$, corresponding to a stiff 
matter behavior.  The scale factor is polynomial: in fact, both points describe 
models with $a\sim t^{1/3}$ in the phase near the Big Bang $a\rightarrow 0$.\\

Also the sinks are three and for their interpretation one follows the same
approach as for the previous ones. The point $\{ 0,1,0 \}$, in the invariant
subset $X_H=1$, is an asymptotically expanding model with exponential scale 
factor $a \sim e^{H_0\, t}$ and parameters $q=w=-1$.  The other two sinks 
$\{ \pm1,-1,\mp\sqrt{3/2} \}$ represent polynomially collapsing models with 
$a\sim t^{-1/3}$ and stiff matter behavior.  All these points describe phases 
with negligible curvature.\\

Finally, the variable space has a number of saddle points which act as transients
for some trajectories.  The points $\{ \pm1/2 , 1 , \pm\sqrt{3/2} \}$ describe 
expanding flat models with scale factor $a \sim t^{4/3}$, while at points
$\{ \pm1/2 , -1 , \mp\sqrt{3/2} \}$ the models are collapsing with scale factor
$a \sim t^{-4/3}$. For both pairs of points we have $w=-1/2$, which is a 
common EoS parameter found in quintessence models.  The points 
$\{ \pm1/\sqrt{3} , \sqrt{3}/2 , \pm\sqrt{3/2} \}$ present a different behavior 
from the other cases, because the curvature here is not negligible with respect 
to the expansion: in this case the model is linearly expanding with $a \sim t$ 
and the EoS parameter $w=-1/3$ is the one corresponding to a ``curvature fluid''.  
The same goes for the points $\{ \pm1/\sqrt{3} , -\sqrt{3}/2 , \mp\sqrt{3/2} \}$, 
with the difference that these describe collapsing phases with $a \sim t^{-1}$.\\

One should note that the system defined by the Hamiltonian of
Secs.~\ref{sec:cosmomod} and \ref{sec:UDMHam} is conservative, in the sense that Liouville's theorem ensures 
preservation of the phase-space volumes under the Hamiltonian flow.  Or, equivalently, there 
are no attractors in the Hamiltonian system when the trajectories are plotted in canonical 
coordinates.  We stress this since it has been shown that apparent attractors can appear when Lagrangian variables are used instead of the canonical ones \cite{remmen2013}.  However, it's a quite well-known procedure to transform a non-conservative system into a conservative one and vice versa by combining variables and evolution parameter transformations \cite{Henrard1991}, which is actually what was performed in Sec.~\ref{sec:UDMboxD}.

 \begin{figure}[!hb]
 \begin{center}
 \includegraphics[width=0.43\textwidth]{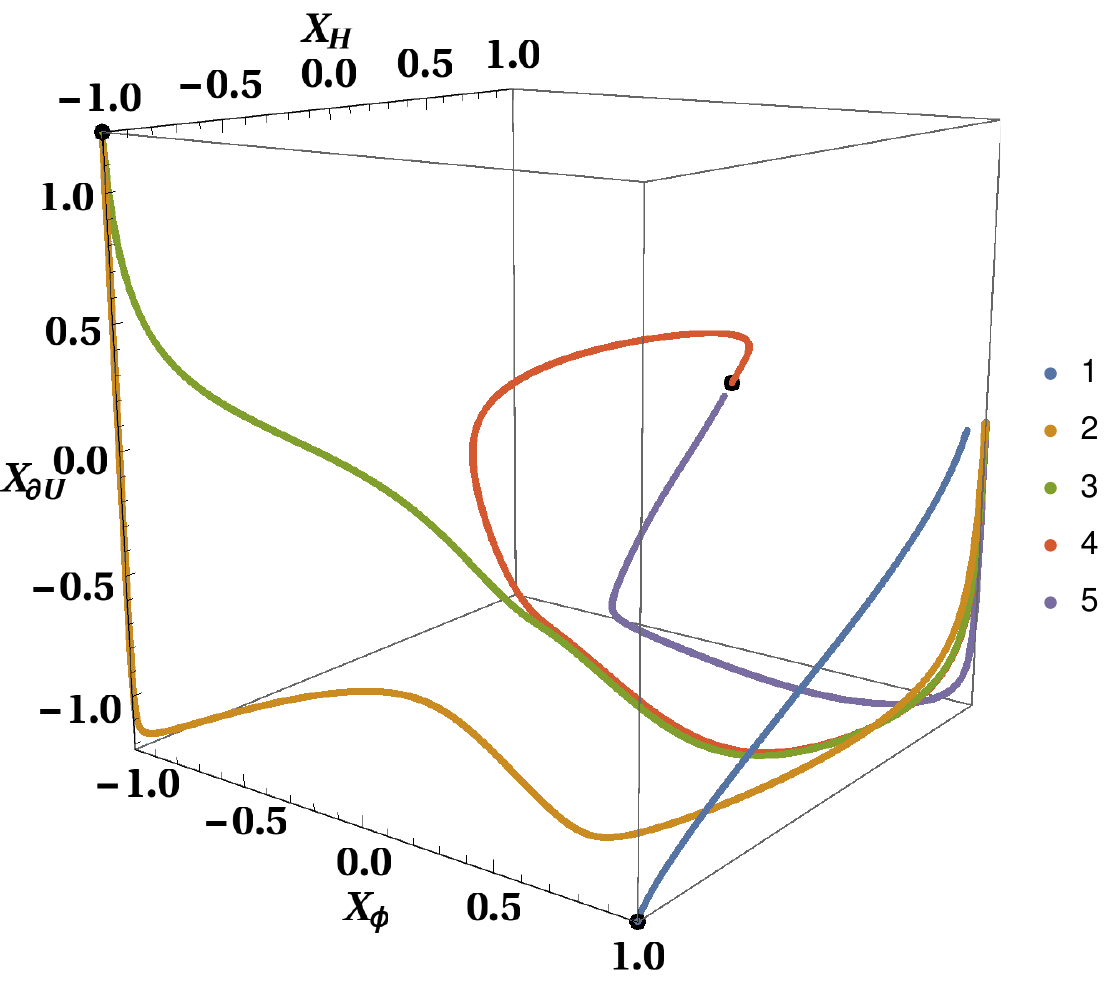}
 \includegraphics[width=0.43\textwidth]{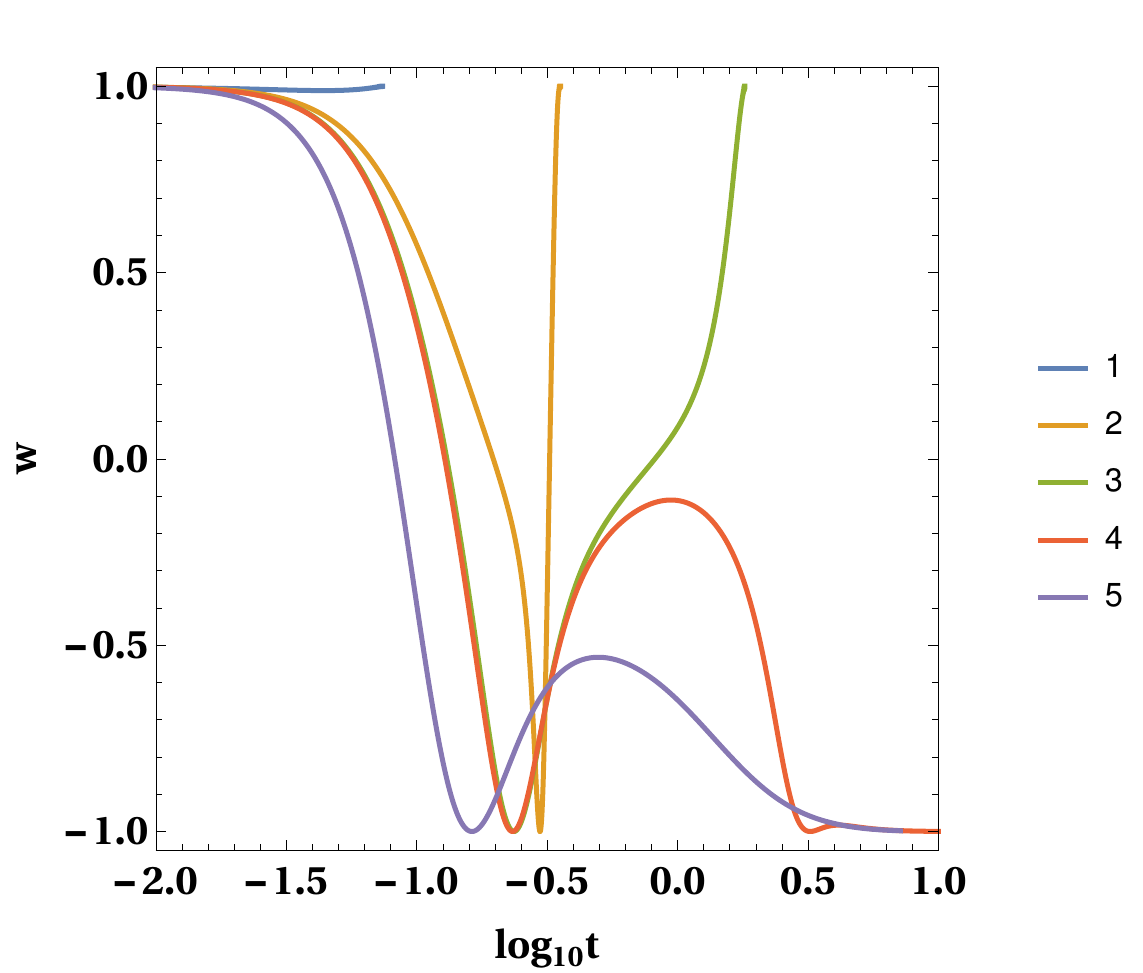}
 \caption{\label{traj} Top panel: five representative trajectories in the 
 variable space; the initial conditions corresponding to the numbers are given
 in the text; the dots indicate the attractors. 
 Bottom panel: the evolution of the effective EoS parameter as a function of
 the cosmic time for the same initial conditions.}
\end{center}
\end{figure}

\section{Orbital analysis} \label{sec:orbital}


The critical points tell us about the basic global structure of the model, but
they do not tell us all the details of how individual initial conditions evolve.
To study individual
evolutions is like doing orbital analysis of the system. In \cite{LGBC08} the
dynamical study was reduced to the case $c_1=c_2=1$ with $k=10^{-3}$ which was
called the ``semi-flat'' model, and in particular the following five initial
conditions were considered:
1) $ y = 0.0001 $, $ x = 0 $, $ p_x = 0.006 $,
2) $ y = 0.0001 $, $ x = 0 $, $ p_x = 0.03 $,
3) $ y = 0.0001 $, $ x = 0 $, $ p_x = 0.04 $,
4) $ y = 0.0001 $, $ x = 0 $, $ p_x = 0.041 $,
5) $ y = 0.0001 $, $ x = 0 $, $ p_x = 0.099 $, 
where the missing variable $ p_y $ was evaluated through eq.~\eqref{Hamxy} with
$ {\cal H} = 0 $ and by choosing the negative sign in front of the square root 
of $ p_y^2 $.

The analysis in Sec.~\ref{sec:critical} showed us that the structure of the parameter
space of UDM cosmology is independent of $\alpha$ in the range $\alpha\geq-\frac{1}{2}$,
i.e. for any pair of $c_1,~c_2$ satisfying such condition the nature of the critical points 
does not change. Thus, for simplicity and without loss of generality one can indeed choose 
$c_1=c_2=1$ to study the dynamics.  

In the top panel of Fig.\ref{traj} we show the 5 representative trajectories
evolving inside the 3-dimensional variable space. We implement the same initial
conditions by translating them to the variables $\{ X_{\phi}, X_H, X_{\partial U} \}$, 
through the following transformation formulas
\begin{align} 
 X_\phi=\frac{y\dot{x}-x\dot{y}}{\sqrt{3^{4/3}k(y^2-x^2)^{4/3}+(y\dot{y}-x\dot{x})^2}} \label{osc2boxXfi} \quad,\\
 X_H=\frac{y\dot{y}-x\dot{x}}{\sqrt{3^{4/3}k(y^2-x^2)^{4/3}+(y\dot{y}-x\dot{x})^2}} \label{osc2boxXH}  \quad,\\
 X_{\partial U}=\sqrt{\frac{3}{2}}\frac{c_1 x y}{(c_1+c_2) y^2-c_2 x^2} \label{osc2boxXdv} \quad.
\end{align}
Orbit 1 ends in the sink $\{ 1 , -1 , -\sqrt{\frac{3}{2}} \}$ following
approximately the flow of the invariant set shown in the top panel of
Fig.~\ref{xphi_inv}. Orbits 2 and 3 end up in the sink $\{ -1 , -1 , \sqrt{\frac{3}{2}} \}$.
These different recollapsing scenarios could not be discerned in the Hamiltonian formalism.
Orbits 4 and 5 end up in the de Sitter sink $\{ 0 , 1 , 0 \}$. Moreover, we know 
now from a dynamical point of view why orbits 3 and 4 split from each other
after evolving almost identically for considerable time. Namely, orbits 3 and 4
follow the flow along the separatrix on the invariant set shown in the right
panel of Fig.~\ref{xdu_inv}, until they reach the saddle point
$(-\frac{1}{\sqrt{3}},\frac{\sqrt{3}}{2},-\sqrt{\frac{3}{2}})$ where they split
following two different eigendirections. The evolution of the EoS parameter for
all these orbits as function of the cosmic time is shown in the bottom panel of
Fig.~\ref{traj}.  Apparently all the initial conditions correspond to an initial
epoch where stiff matter EoS $w=1$ dominates. Orbit 1 doesn't deviate much from
this behavior and quite soon enters the recollapsing phase. All the other orbits
passing through all the intermediate values of $w$ reach a transitory phase of
exponential expansion ($w=-1$). After this common de Sitter phase, orbits 2 and
3 return to a stiff-matter phase by following a recollapsing dynamics.  For
orbit 4 instead the EoS parameter momentarily increases up to a dust-like
behavior $w=0$, before returning to the de Sitter expansion with $w=-1$. As for
orbit 5, its final fate is the same as trajectory 4 but the intermediate phase
reaches $w=-1/2$.  

\begin{figure}[ht!!!]
 \begin{center}
 \includegraphics[width=0.4\textwidth]{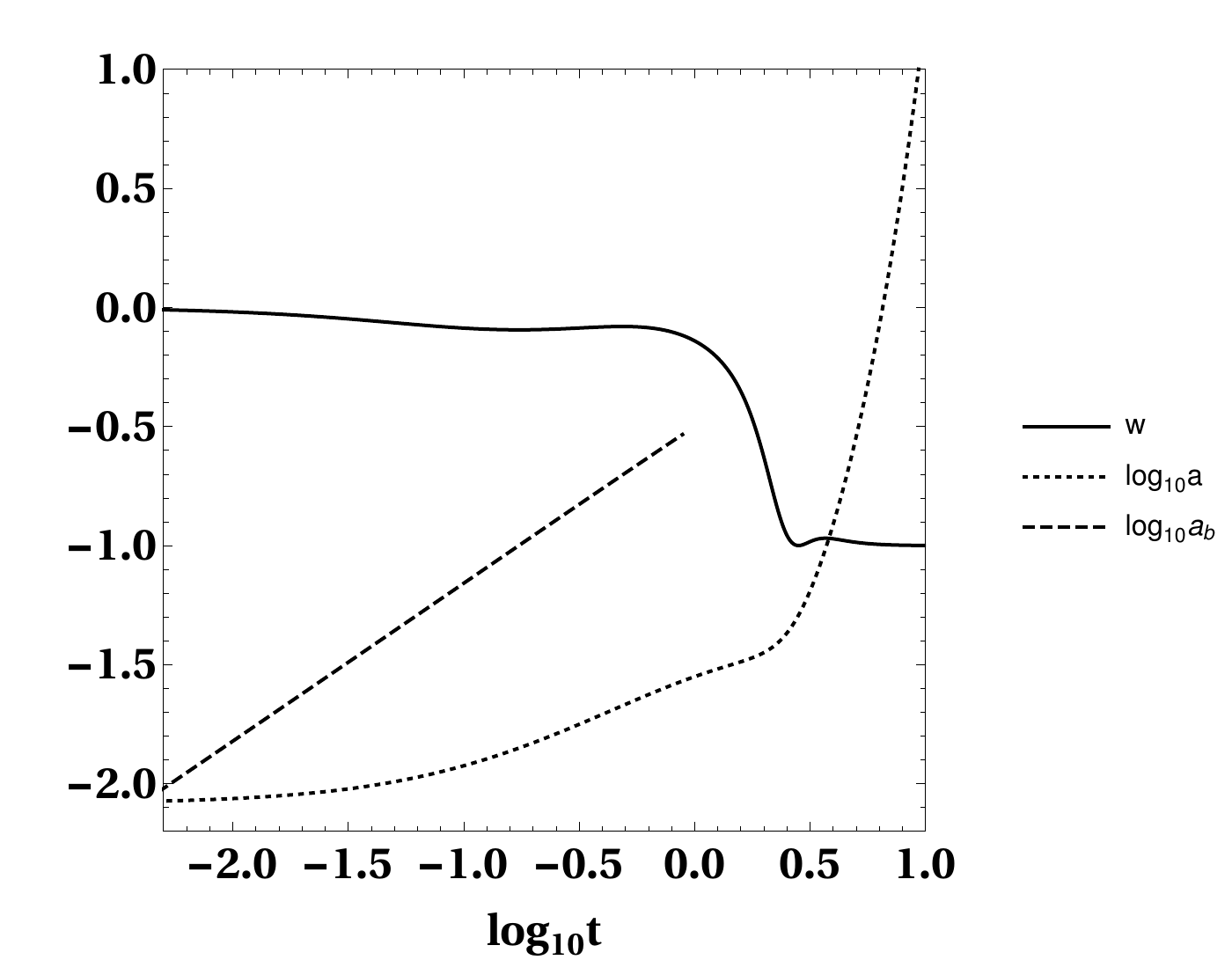}
 \caption{\label{Comwat} The scale factor $a$  and the EoS parameter $w$ as
 functions of the cosmic time $t$ for a trajectory starting from
 $(-1/\sqrt{2},0.8,-\sqrt{3/2}+0.03)$. $a_b$ shows the evolution of the scale
 factor in the matter dominated era, if we assume constant $w$. 
 }
\end{center}
\end{figure}

\subsection{Remarks on the equation of state}\label{sec:RemEoS}

The evolution of the scale factor of orbits 4 and 5 should be qualitatively close
to the $\Lambda$CDM model \cite{LGBC08}. However, by comparing Fig.~1(b) of
\cite{LGBC08} with the bottom panel of Fig.~\ref{traj}, we can see that the EoS 
parameter values are not in complete agreement with the inclinations of the scale 
factor's logarithmic plot. For example the scale factor of orbit 4 appears to
follow the matter dominated era inclination, i.e. $a\propto t^{2/3}$, at
$-0.6\lesssim \log_{10} t \lesssim 0$, but for this time range $w$ undergoes a
transition from $-1$ to $0$ and back to $-1$. Thus, the following two questions arise:
a) are the UDM models compatible with an effective description using a barotropic
equation of state? b) Can one infer the dominating era of a component
of the universe by the evolution of the scale factor?

To answer these questions we searched for a trajectory having $w\approx 0$ for a
considerable time. Since we want to impose $w=0$, from eq.~\eqref{wEoS} we get
that this is achieved when $X_\phi^2=1/2$. Moreover, we want the scenario to be
expanding and the curvature to be negligible in comparison to the expansion, so
$X_H\approx1$. In order to find the appropriate value for $X_{\partial U}$ we
performed a scan in the range of its possible values, i.e. 
$[-\sqrt{3/2},\sqrt{3/2}]$, and Fig.~\ref{Comwat} shows the scenario which we were
looking for. Namely, the parameter $w$ (continuous curve) behaves like dust from
the start until $t \approx 1$, with negligible variations; in this time range,
since the fluid is barotropic, one would anticipate that the scale factor should
grow approximately like $a\propto t^{2/3}$ since the scale factor for a
barotropic fluid grows like
\begin{align}
 a_b\propto t^{\frac{2}{3(1+w)}} \quad, \label{abarot}
\end{align}
when $w$ is almost constant. The dashed line shows $a_b$ when $w$ is given by 
eq.~\eqref{wEoS}, with the initial value of $a_b$ appropriately rescaled in order to 
fit the initial value of the dotted line. 
The dotted line, in turn, shows the evolution of the scale factor $a$ as it
comes from the equations of motion~\eqref{phip}-\eqref{dup} when the inversion
equation~\eqref{scfnewvar} is applied; note that in the latter case no assumption
is made regarding the EoS. We see that the dashed and the dotted curves do not 
match, so one would assume that the scalar field is not behaving like a barotropic
fluid in the matter dominated era. However, such conclusion would be wrong. 
If one  wants to do the check properly, then from the stress energy conservation, i.e.
\begin{align}
 \dot{\rho}_\phi+3~H(\rho_\phi+p_\phi)=0 \quad,
\end{align}
one gets, by assuming a barotropic fluid behavior for the scalar field, that 
\begin{align} \label{densityEoS}
 \rho_\phi(a)=\rho_\phi(a_0)\exp{\left(-\int_{a}^{a_0}\frac{3(1+w(\tilde{a}))}{\tilde{a}}d\tilde{a} \right)} \quad.
\end{align}
Eq.~\ref{densityEoS} is then substituted in eq.~\eqref{efe1}. The solution of the resulting equation
after the above substitution is giving the scale factor with respect to the
cosmic time. Note that we take the positive root of $\dot{a}$, since we
anticipate an expanding universe. By following the above procedure in which we
employ the numerical found data one gets a scale factor evolution which
reproduces the dotted line of Fig.~\ref{Comwat}. Thus, a) the UDM scalar field
is behaving like a barotropic fluid and b) we cannot infer the dominating era of
a component of the universe by the evolution of the scale factor, i.e.
eq.~\eqref{abarot} is not proper for the present discussion.

\begin{figure}[h!!!]
 \begin{center}
 \includegraphics[width=0.43\textwidth]{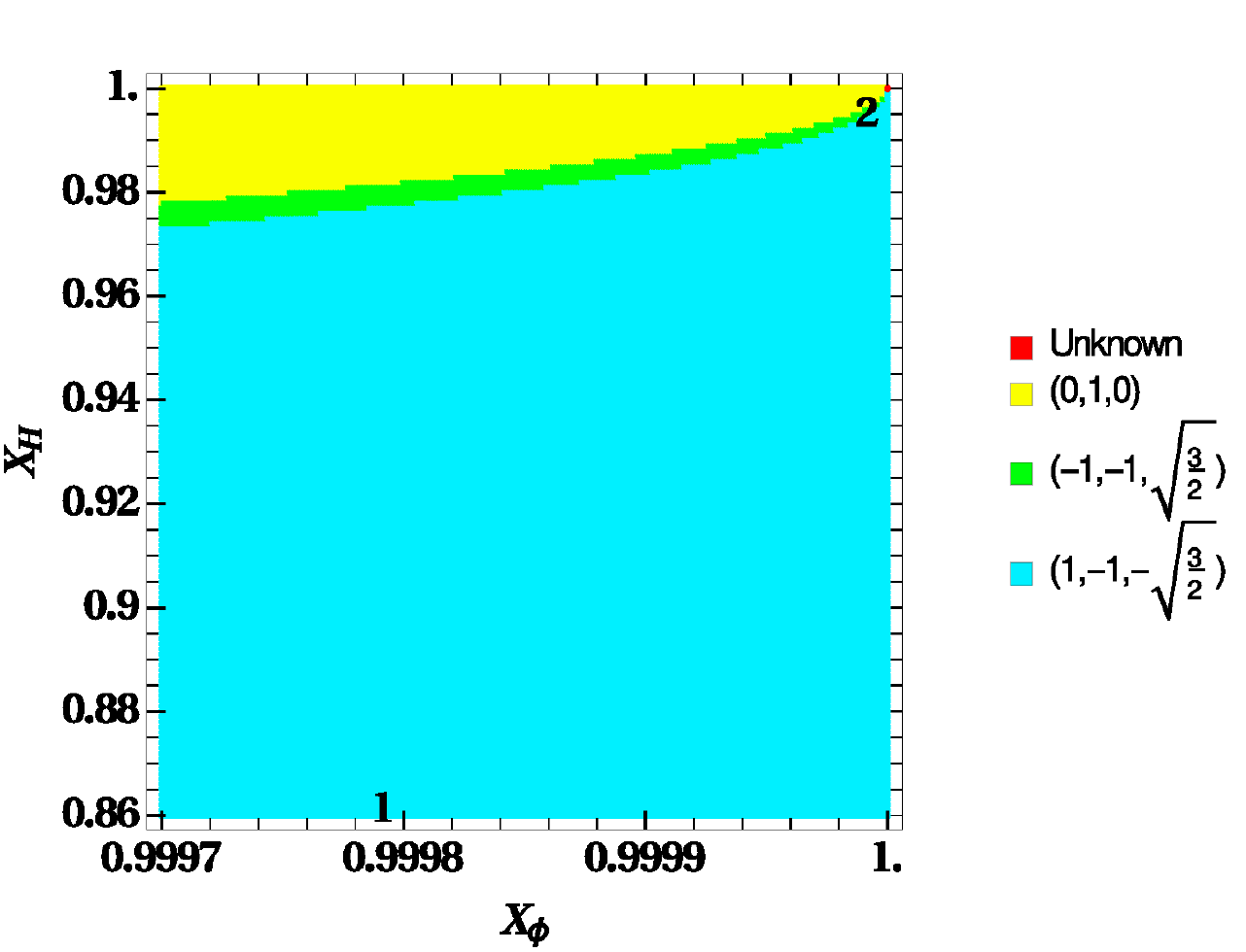}
 \caption{\label{grid12} A portion of the plane $X_{\partial U}=0$. The numbers
 1 and 2 identify the initial conditions for the respective trajectories of
 Fig.\ref{traj}.  The colors correspond to the basins of attraction of the three
 sinks for the points in the plane.}
\end{center}
\end{figure}

\subsection{Remarks on the chaotic behavior}

In the picture of the two coupled oscillators (Sec.~\ref{sec:UDMHam}) the
dynamics indicated chaotic scattering, which would mean that the basins of
attraction in the UDM box description should have fractal structure. 
The basins of attraction are defined by the sink a trajectory ends up.  

\begin{figure*}[ht!!!]
 \begin{center}
 \includegraphics[width=0.3\textwidth]{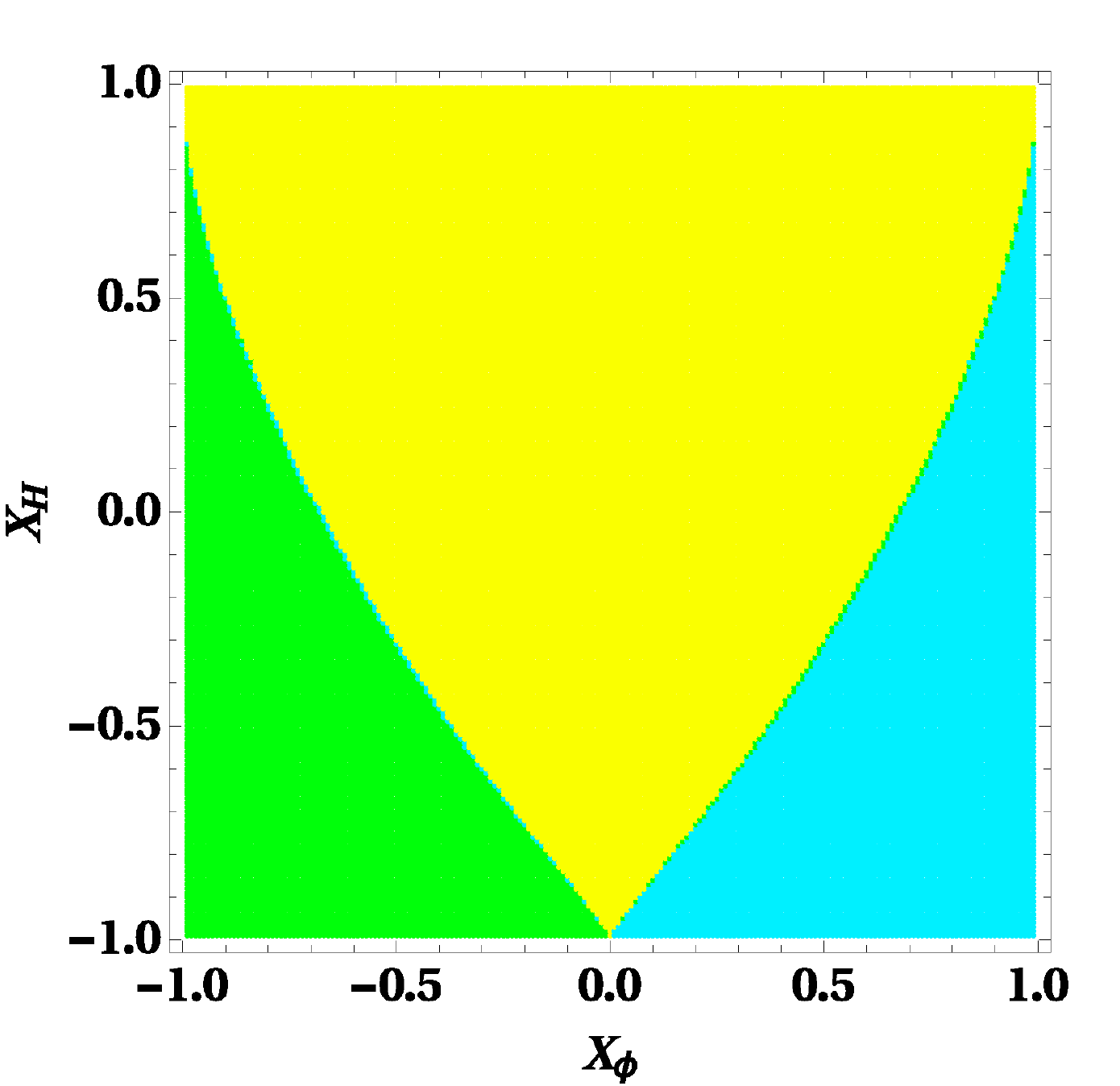}
 \includegraphics[width=0.3\textwidth]{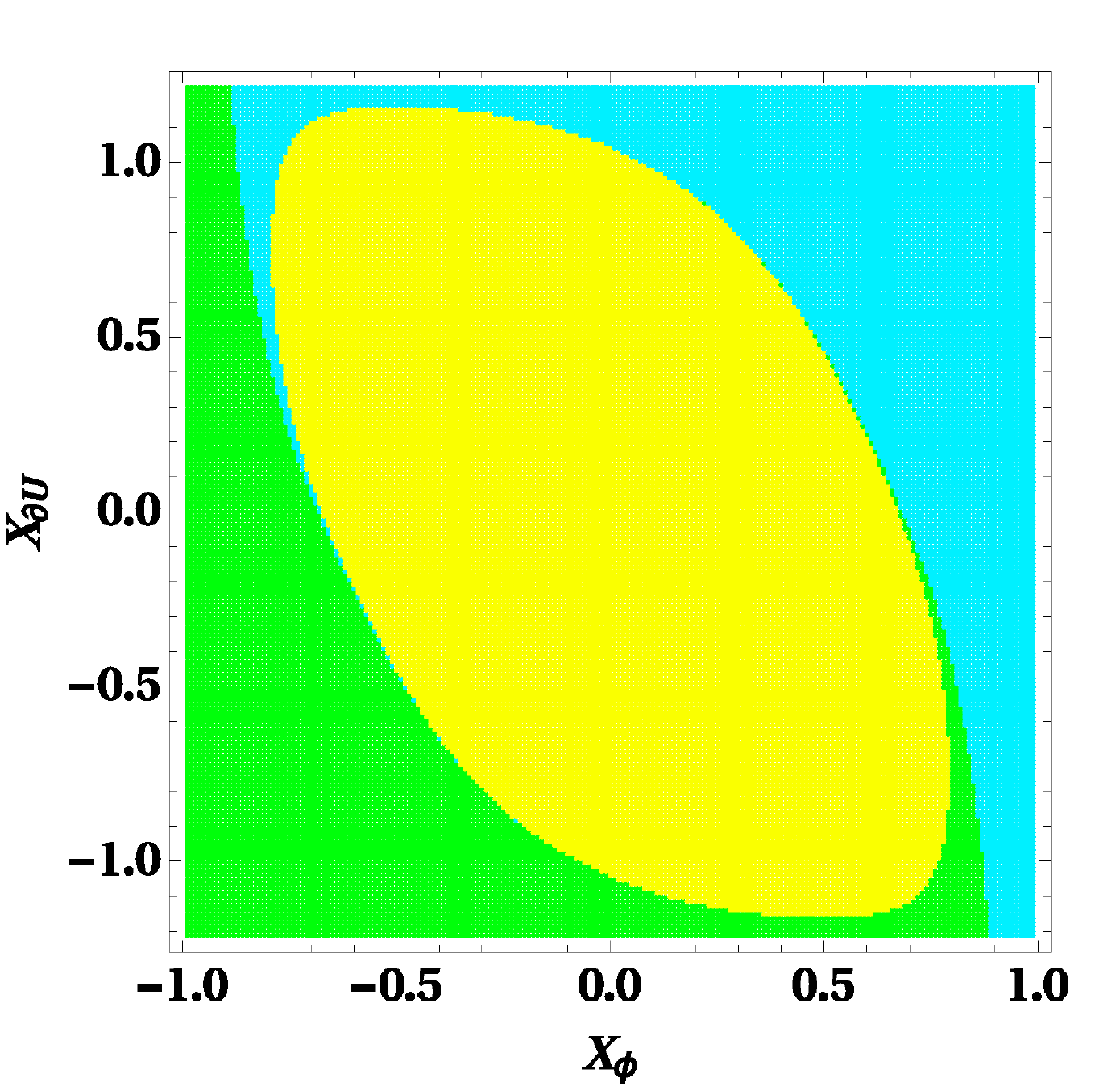}
 \includegraphics[width=0.3\textwidth]{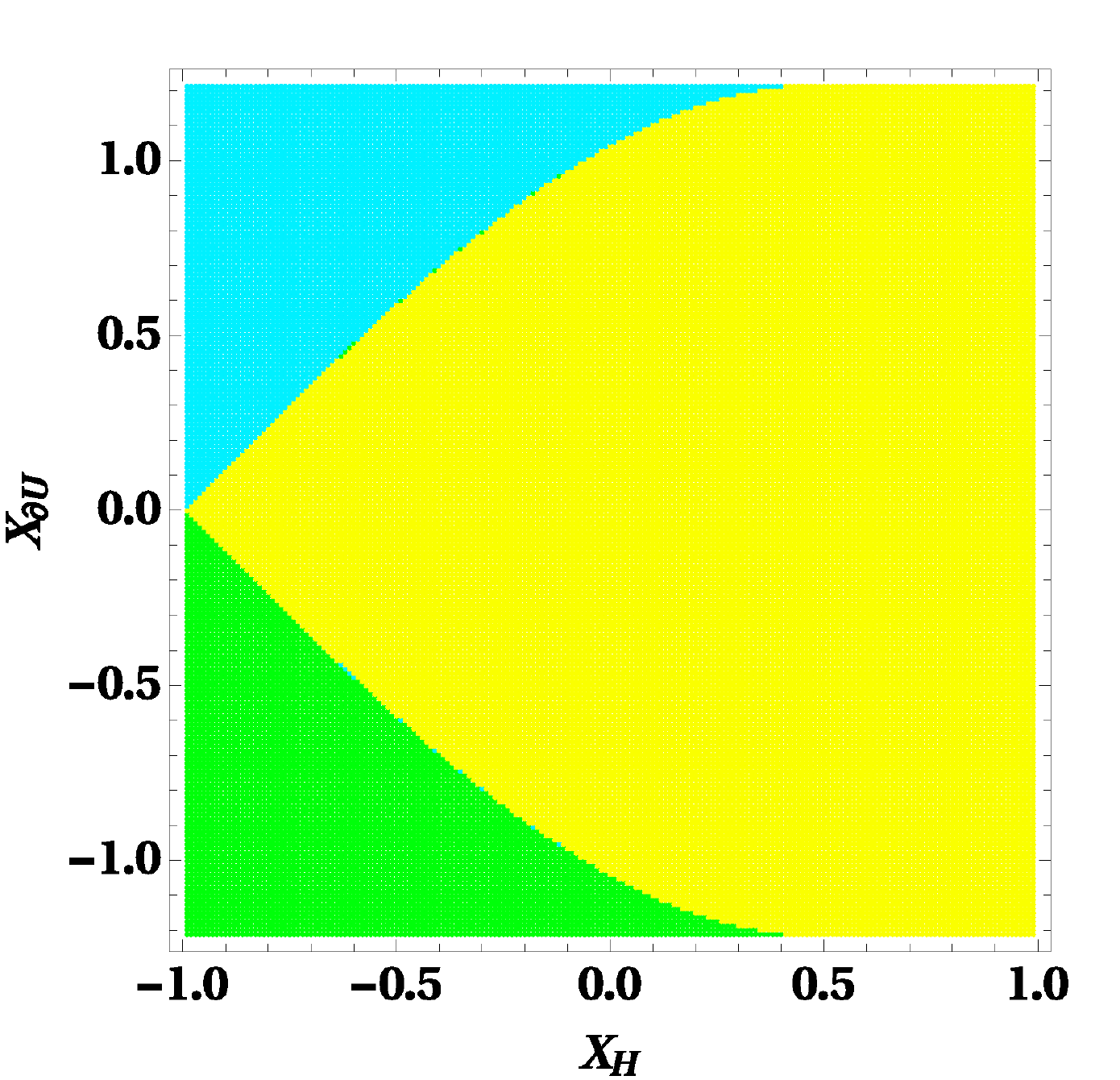}
 \caption{\label{grids} The basins of attraction of the three sinks for the
 points in the planes $X_{\partial U}=0$ (left panel), $X_H=0$ (central panel)
 and $X_{\phi}=0$ (right panel). The colors correspond to the basins of attraction of the three
 sinks indicated in Fig.\ref{grid12}.}
\end{center}
\end{figure*}
We start our analysis in the parameter range in which the five representative
trajectories were chosen. Since all the five orbits start from $x=0$,
the transformation formula~\eqref{osc2boxXdv} gives that the plane
on which they all lie is $X_{\partial U}=0$. Fig.~\ref{grid12} shows the
basins of attraction on the latter plane; the position of orbit 1 and 2 is
indicated by the respective numbers. Actually, number 2 also indicates the area
where 3,~4,~5 approximately lie. The colors denoting the basins of each sink in
Fig.~\ref{grid12} are smoothly separated with no evidence of smaller structures
or mixing. Thus, there is no indication of fractal structure, which in turn
implies absence of chaos. The absence of fractality was confirmed by choosing
several different planes; a sample of them is shown in Fig.~\ref{grids}.  This result is
in agreement with the analysis of \cite{topo}.  In particular, using the notation of \cite{topo}, the condition
for chaoticity is $\phi_0>4 m_p/\sqrt{3}$: since $2\phi_0=c^{-1}$ and 
$m_p^2=8\pi$ in our convention, our parameter $c$ is not in the chaoticity range.

\section{Conclusions} \label{sec:concl}

We have analyzed a Unified Dark Matter model in the background of a 
Friedmann-Robertson-Walker spacetime by both revisiting the study performed in 
\cite{LGBC08} and providing new results. A specific variable transformation has
allowed us to bound in a box the UDM model in the case of FRW metric with
positive spatial curvature. In this new set of variables the critical points of
the dynamical system, i.e. sinks, sources and saddle points, were found and
interpreted according to their dynamical and cosmological features. In particular,
we have found that for the UDM models there are two dynamically distinct (but
cosmologically indistinguishable) recollapsing scenarios, a feature that was not
found in the previous study \cite{LGBC08}, and that at the vicinity of
some of the saddle points the scalar field is behaving like a curvature fluid. 

Characteristic evolution scenarios discussed in \cite{LGBC08} were revisited and
analyzed in the framework of the new variables. It was found that these scenarios
start from a stiff matter equation of state $w=1$, and most of them reach the 
de Sitter phase $w=-1$ by a smooth transition through all the values of $w$ in 
between. At the de Sitter phase either the scenarios keep exponentially growing
after a transient phase resembling a recollapse, or they actually recollapse.
By taking advantage of the better insight provided by the new variables, we 
discussed the barotropic character of the UDM scalar field. While it makes sense
to characterize {\it a critical point} by associating it to 
a constant effective EoS parameter $w$ through eq.~\eqref{wEoS}, different 
$w$-dominated epochs {\it along an orbit} are not described by the respective 
scale factors given by eq.\eqref{abarot} with constant EoS parameter: one has 
to keep in mind that the whole integrated effect along the orbit has to be considered,
i.e. one has to employ eq.\eqref{densityEoS}.

Lastly, by studying the basins of the sinks in the variable space, we did not
find any sign of fractal structure.  This means that the system, for the range of parameters considered, is not chaotic, contrary to the findings of \cite{LGBC08} but in agreement with the analysis of \cite{topo}.
This contradiction might resemble the mixmaster case, where contradicting indications have resulted in a long debate on whether that system
was chaotic or not, see e.g. \cite{Cornish97,Contop99} and references therein.
However, as was mentioned in the introduction, Lyapunov-like indicators
are not completely reliable tools for detecting chaos in unbounded systems.  So, given the current results and
the investigation of \cite{topo}, we can claim that the findings of \cite{LGBC08} based on Lyapunov-like indicators were misleading.

\begin{acknowledgments}
G.A. is funded by the grant GACR-14-37086G of the Czech Science Foundation.
G.L.-G. is supported by UNCE-204020.  
\end{acknowledgments}

\appendix

\section{Alternative form of potential}\label{appen}

Let's consider here the case in which the potential is given by
\begin{align}
 U(\phi) &= b_1 \left(1+\cosh\left( \sqrt{\frac{3}{8}}\, \phi \right)\right)^3 \quad,
\end{align}
where $b_1\geq0$ in order to have non-negative mass for the scalar field.
The dynamical system can be constructed in a completely analogous way as
for the case of potential eq.\eqref{pot}, the only difference being that in this 
case the variable $X_{\partial U}$ is bounded in the interval 
$\left[ -(3/2)^{3/2} , (3/2)^{3/2} \right]$.  This has repercussion on the
location of some critical points of the system and their cosmological
interpretation, but their total number and their stability character are unchanged 
with respect to the system analyzed in detail in the main text.
For completeness, in table \ref{tab2} we list the critical points and their
features.  We can immediately see that the only cosmological difference
regards the behavior of the model in the saddle points with coordinates
$X_{\phi}=\pm3/4$.  By assuming the barotropic equation of state $w=1/8$,
the effective scale factor would be $a\sim t^{16/27}$, which is
in between radiation-like and dust-like behavior.

\setlength{\extrarowheight}{10pt}
\setlength\tabcolsep{10pt}
\begin{table}
  \begin{tabular}{c||c|c|c}
    $ \{ X_{\phi} , X_H , X_{\partial U} \}$ & \ stability & $q$ & $w$ \\\hline
    $ \{ 0 , -1 , 0 \}$ & source & -1 & -1 \\
    $ \{ 0 , 1 , 0 \}$ & sink & -1 & -1 \\
    $ \{ \pm1 , 1 , \pm\left(\frac{3}{2}\right)^{3/2} \}$ & source & 2 & 1 \\
    $\{ \pm1 , -1 , \mp\left(\frac{3}{2}\right)^{3/2} \}$ & sink & 2 & 1 \\
    $\{ \pm1 , 1 , \mp\left(\frac{3}{2}\right)^{3/2} \}$ & saddle & 2 & 1 \\
    $\{ \pm1 , -1 , \pm\left(\frac{3}{2}\right)^{3/2} \}$ & saddle & 2 & 1 \\
    $\{ \pm3/4 , 1 , \pm\left(\frac{3}{2}\right)^{3/2} \}$ & saddle & $\frac{11}{16}$ & 1/8 \\
    $\{ \pm3/4 , -1 , \mp\left(\frac{3}{2}\right)^{3/2} \}$ & saddle & $\frac{11}{16}$ & 1/8 \\
    $\{ \pm\frac{1}{\sqrt{3}} ,\, \pm\frac{3\sqrt{3}}{4} ,\, \left(\frac{3}{2}\right)^{3/2} \}$ & saddle & 0 & -1/3 \\
    $\{ \pm\frac{1}{\sqrt{3}} ,\, \mp\frac{3\sqrt{3}}{4} ,\, -\left(\frac{3}{2}\right)^{3/2} \}$ & saddle & 0 & -1/3 \\
  \end{tabular}
\caption{\label{tab2} Critical points of the system with potential given by eq.\eqref{pot2}, their stability and 
corresponding cosmological parameters.}
\end{table}

\end{document}